\documentclass[11pt,a4paper]{article}
\pdfoutput=1
\usepackage{jcappub}
\usepackage{bm}
\newcommand{\Mpl}{M_{\rm pl}}

\newcommand{\fnl}{f_{\mathrm{NL}}}

\newcommand{\be}{\begin{equation}}
\newcommand{\ee}{\end{equation}}
\newcommand{\bea}{\begin{eqnarray}}
\newcommand{\eea}{\end{eqnarray}}

\newcommand{\di}{ {\rm d}}

\graphicspath{{./}}
\notoc

\begin{document}

\title{Exploring Two-Field Inflation in the Wess-Zumino Model}

\author{John Ellis$^1$,}
\author{Nick E. Mavromatos$^1$,}
\affiliation{\vspace{2mm}
$^1$ Theoretical Particle Physics and Cosmology Group, Department of Physics, King's College London, London WC2R 2LS, UK; \\
Theory Division, Physics Department, CERN, CH1211 Gen\`eve 23, Switzerland}

\author{David J. Mulryne$^2$ }
\affiliation{\vspace{2mm}
$^2$ School of Physics and Astronomy, Queen Mary
University of London, Mile End Road, London, E1 4NS, UK}

\emailAdd{John.Ellis@cern.ch, \\ ~~~~~~~~~~~Nikolaos.Mavromatos@kcl.ac.uk, \\ ~~~~~~~~~~~d.mulryne@qmul.ac.uk}

\abstract{We explore inflation via the effective potential of the minimal Wess-Zumino model,
considering both the real and imaginary components of the complex field. Using transport 
techniques, we calculate the full allowed range 
of $n_s$, $r$ and $\fnl$ for different choices of the single free parameter, $v$, 
and present the 
probability distribution of these signatures given a simple choice for the prior distribution 
of initial conditions. 
Our work provides a case study of multi-field inflation in a simple but realistic 
setting, with important lessons 
that are likely to apply more generally. 
For example, we find that there are initial conditions consistent with observations of $n_s$ and $r$
for values of $v$ that would be excluded if only evolutions in the real field direction were 
to be considered, and that these may yield enhanced values of $\fnl$. 
Moreover, we find that initial conditions fixed at high energy density, where the potential 
is close to quartic in form, can still lead to 
evolutions  in a concave region of the potential during the observable 
number of e-folds, as preferred by present data. The Wess-Zumino model therefore
provides an illustration that multi-field dynamics must be 
taken into account when seeking to understand fully the 
phenomenology of such models of inflation.\\
~\\
KCL-PH-TH/2014-01, LCTS/2014-01, CERN-PH-TH/2014-003 \\
}

\maketitle

\section{Introduction}

Data from the Planck satellite \cite{Ade:2013zuv, Ade:2013uln, Ade:2013ydc} provide important new constraints on models of
cosmological inflation, restricting the allowed values of the scalar spectral index, $n_s$,
the tensor-to-scalar ratio, $r$, and the non-Gaussianity parameter, $\fnl$.
The $n_s$ and $r$ constraints put pressure on many single-field models of
inflation, particularly those with monomial potentials of the form $\phi^n$. Indeed,  even
the predictions of the simple $\phi^2$ chaotic inflation model lie outside the $(n_s, r)$ region
favored by the Planck data at the 68\% confidence level. On the other hand, 
there is no hint that $|\fnl|$ is significantly different from zero, as predicted in some multi-field scenarios.

The constraints on $n_s$ and $r$ motivate the reconsideration of non-monomial single-field
models with concave regions, such as models  of the form   
$\phi^2(v - \phi)^2$ that include a hilltop. It
was shown in~\cite{Croon:2013ana} that this model yields Planck-compatible 
inflation for suitable values of $v \gg \Mpl$ and  initial field values $\phi \sim v/4$. It was also
pointed out that this potential could be interpreted as the restriction to
real field values in the minimal Wess-Zumino
model with superpotential $W = (\mu/2) \Phi^2 - (\lambda/3) \Phi^3$ and scalar potential
$V = |\partial W/\partial \Phi|^2$. Study of this model is well-motivated both theoretically, since it
is the simplest supersymmetric model, and cosmologically, since it yields predictions for
$n_s$ and $r$ that are compatible with the Planck measurements.
Another interesting feature of the model  is that it might provide a viable
extension of the minimal supersymmetric seesaw model of neutrino masses, 
if one interprets $\Phi$ as a right-handed singlet neutrino superfield. In this case, one
could envisage a scenario of chaotic sneutrino inflation, followed by leptogenesis during the
subsequent reheating~\cite{Ellis:2003sq}.

This Wess-Zumino model, however, necessarily yields a two-field inflationary potential, since both 
the real and imaginary components of the complex field $\Phi$ are 
light at horizon crossing. In addition to increasing the possible range of predictions
for $n_s$ and $r$, such a two-field model may also yield larger values of $\fnl$. Therefore, in this paper we 
explore the cosmological phenomenology of the different inflationary trajectories allowed  
in this two-field setting, fully accounting for the presence of isocurvature modes. 
In particular, we extend the earlier study~\cite{Croon:2013ana} to explore the full range of initial conditions 
that are consistent with Planck data when the evolution of both components of the
Wess-Zumino field are considered, finding some examples of initial conditions that yield
values of $n_s$ and $r$ similar to the Starobinsky model of $R^2$ inflation~\cite{starobinsky}.
The large range of possible initial conditions in such a two-field 
model motivates a statistical study, in which 
a probability distribution of the observational signatures is calculated. We perform this calculation 
for a simple choice of prior on the parameter space of possible initial conditions, namely a uniform
prior along a contour of constant potential energy density.

The Wess-Zumino model is sufficiently simple a setting for an extensive study of these issues. However, it 
is also sufficiently realistic that it serves as an instructive example of the changes that may occur when 
the restriction to single-field inflation is relaxed in models with a complex field, or in models 
that contain other additional light scalars. 
Such models are generic when inflation is embedded in theories of particle physics  
beyond the standard model, so confronting
multi-field inflation with observations is essential for understanding
whether such models are consistent with current data. In particular, spin-zero
fields are always paired in supersymmetric models. 

In this paper, our approach to calculating observable signatures 
will necessarily be a numerical one. Analytical 
calculations of $n_s$, $r$ and $\fnl$ are only possible for models in which slow-roll is an extremely good 
approximation for the entire observable inflationary phase and moreover for which the 
potential is of product or sum-separable form \cite{ Vernizzi:2006ve, Choi:2007su}, which is 
rarely the case in realistic examples (see e.g.\cite{Dias:2012nf, McAllister:2012am, Nolde:2013bha} 
for recent 
numerical studies with similarities to ours).

The paper is structured as follows. In \S~\ref{calc} we briefly review the techniques we 
use to calculate observables in multi-field inflation -- readers familiar with this material, or who 
would rather go straight to the model and results, can skip this section. In \S~\ref{model} we introduce the 
minimal Wess-Zumino model. In section \S~\ref{results}  we calculate the 
full range of observational signatures in this model, as well as their statistics, and we conclude in \S~\ref{conclusions}.

\section{Calculational Techniques}
\label{calc}

In order to calculate the parameters $n_s$ and $r$, we use perturbation theory and simple quantum field theory (QFT)
techniques to numerically evolve the equal-time two-point correlation functions of scalar field 
perturbations in Fourier space. We choose Bunch-Davis vacuum initial conditions, and evolve the correlation functions 
until the end of inflation (this choice is discussed further below).  
These can then be related to the statistics of the uniform density curvature 
perturbation, $\zeta$ \cite{Malik:2008im}, which is directly constrained by observation. 
In the case of the three-point function, which is required to calculate the local 
non-Gaussianity parameter $\fnl$, 
we assume for simplicity that the 
field fluctuations are Gaussian at the point where the 
Fourier modes of interest cross the cosmological horizon, which is a good 
approximation for canonical models~\cite{Seery:2005gb}, and employ  
second-order perturbation equations on super-horizon scales to 
evolve the scalar field three-point function to the end of inflation. This approximation is
equivalent to the oft-used ``$\delta N$" \cite{Lyth:2005fi} approach to calculating $\fnl$, and is likely to 
be sufficiently accurate for the calculation of $\fnl$ in this model~\footnote{However, we note that
work is now in progress to develop the techniques necessary to 
derive more precise predictions for the bispectrum~\cite{us}, as may be
required  in multi-field models that yield larger values of $\fnl$ than the Wess-Zumino model studied here,
which could be observable in the foreseeable future.}.

The techniques can be summarized as follows. 
Consider a canonical set of fields $\varphi_a$, where $a$ runs from $1$ to $n$. 
At linear order in perturbation theory, perturbations in these fields on flat slices 
of spacetime evolve according to the coupled second-order ordinary differential equations
(ODEs)~\cite{Sasaki:1995aw}
\be
\label{pertEvolve}
\delta \ddot \varphi_a(\mathbf{k}) = -3 H \delta \dot \varphi_a(\mathbf{k}) - \frac{k^2}{a^2} \delta  \varphi_a(\mathbf{k}) - M^2_{ a b} \delta \varphi_b(\mathbf{k})\,,
\ee
where
\be
M^2_{a b} = V_{,ab} - \frac{1}{a^3}\frac{{\rm d}}{{\rm d} t}\left ( \frac{a^3}{H} \dot\varphi_a\dot \varphi_b  \right )\,.
\ee
Introducing $x_\alpha = \{\delta \varphi_a, \delta \dot \varphi_b\}$,  
we can write these equations in a compact form as 
the set of first-order ODEs 
\be
\label{x1Evolve}
\dot x_\alpha(\mathbf{k}) = u_{\alpha \beta}(k) x_\beta(\mathbf{k})\,, 
\ee
where Greek indices now run over all fields and field momenta, and $u_{\alpha \beta}(k)$ can 
be determined in a straightforward way from (\ref{pertEvolve}).

For $k>aH$ the field perturbations cannot be treated classically, and we set up a QFT description by promoting 
$\delta \varphi_a$ to an operator.
Defining the two-point correlation function 
\be
\langle x_\alpha(\mathbf{k}) x_\beta(\mathbf{k'}) \rangle = (2 \pi)^3 \delta(\mathbf{k}-\mathbf{k'}) \Sigma_{\alpha \beta}(t, k)\,,  
\ee 
the Bunch-Davis vacuum initial conditions are such that for $k \ll aH$ the fields are 
uncorrelated and the real part of the two-point function is given by 
\be \Sigma^{r}_{\alpha \beta}(k,t_{\rm init}) = \left| \begin{array}{cc} F_{ab} & C_{a b} \\ C_{a b} & P_{a b} \end{array} \right| \, ,
\ee
where $F_{a b} =  \frac{ H^2}{2 k (a H )^2 } \delta_{a b}$ is the field-field part of $\Sigma^{r}_{\alpha \beta}(t_{\rm init})$, $C_{ab}= -H F_{ab} $ is the initial field-momenta part, and $P_{ab}=  \frac{k^2}{a^2} F_{ab} $ 
the momenta-momenta part. The imaginary part decays on super-horizon scales.


The matrix $\Sigma^r_{\alpha \beta}(k)$ evolves according to the transport equation~\cite{Mulryne:2009kh,Dias:2012nf, Mulryne:2013uka}
\be
\label{transport}
\frac{{\rm d} \Sigma^{r}_{\alpha \beta}(t,k)}{{\rm d} t} = u_{\alpha \gamma}(k) \Sigma^{r}_{\gamma \beta}(t,k) + u_{\alpha \gamma}(k) \Sigma^{r}_{\beta \gamma}(t,k)\,,
\ee
where we recall that $u_{\alpha \beta}$ is defined above through Eqs.~(\ref{pertEvolve}) and (\ref{x1Evolve}). That is, the 
evolution equation for the two-point function of the perturbations follows directly from the evolution equation for the perturbations themselves. 
Evolving the two-point function directly from the Bunch-Davis vacuum has some advantages
when compared with alternatives, such as evolving the mode 
matrices of the QFT \cite{Salopek:1988qh,Huston:2012dv, Easther:2013rva}, or using approximate 
methods for the two-point function such as ``$\delta N$" techniques.
First, it evolves the physical object (the correlation function) directly. 
Secondly, in contrast to the mode matrices themselves, $\Sigma^{r}_{\alpha \beta}(t,k)$ is a real 
valued matrix and is not a highly oscillatory function of time.
Finally, the method can be generalized to higher-order statistics such as the bispectrum, 
allowing simple evolution equations to be written down for the amplitude of 
higher-order correlation functions~\cite{Mulryne:2009kh, Mulryne:2013uka}.

The three-point function of field and field momenta perturbations is defined as
\be
\langle x_\alpha(\mathbf{k}) x_\beta(\mathbf{k'}) x_\gamma(\mathbf{k''})\rangle = (2 \pi)^3 \delta(\mathbf{k}+\mathbf{k'}+\mathbf{k''}) \alpha_{\alpha \beta \gamma}(t, k, k',k'')\, ,   
\ee 
and $\alpha_{\alpha \beta \gamma}$ evolves on super-horizon scales according to the transport equation
\begin{eqnarray}
\label{transport2}
\frac{{\rm d} \alpha_{\alpha \beta \gamma}(t, k,k',k'')} {{\rm d} t} &=& u_{\alpha \mu \nu}(t) \Sigma^r_{\mu \beta}(t,k') \Sigma^r_{\nu \gamma}(t,k'') + u_{\alpha \mu}(t) \alpha_{\mu \beta \gamma}(t,k,k',k'') \nonumber 
\\&+&  {\rm two~cyclic~perms~ \alpha \rightarrow \beta \rightarrow \gamma }\,,
\end{eqnarray}
where the $k$ values associated with 
the free indices are interchanged under the cyclic permutations, and the $u_{\alpha \beta \gamma}$ matrix can be read from the second-order part of the the super-horizon 
evolution equation for $\delta \varphi_\alpha$ (a summary of the 
$u$ matrices is given in Appendix A). Since we restrict
our attention to the local non-Gaussianity parameter $\fnl$, 
we 
will only need to solve this equation from horizon crossing for one $k$ configuration 
in the equilateral limit, $k=k'=k''$.
 
Once $\Sigma$ (hereafter we drop the superscript $r$) and $\alpha$ are calculated at the time of 
interest (for us the end of inflation), these field-space statistics should be converted into the statistics of $\zeta$. 
This is done using the expressions
\be
\label{zeta1}
\Sigma_\zeta(t,k) = N_{\alpha} N_{\beta} \Sigma_{\alpha \beta}(k)\,,   
\ee
and 
\be
\label{zeta2}
\alpha_\zeta(t,k,k,k) = N_{\alpha} N_{\beta} N_{\gamma}(t) \alpha_{\alpha \beta \gamma}(k,k,k) + N_{\alpha \beta} N_{\mu} N_\nu \Sigma_{\alpha \nu}(k) \Sigma_{\beta \mu}(k)   \,,
\ee
where $N_\alpha(t)$ and $N_{\alpha \beta}(t)$ are just functions of background quantities, also 
summarized in Appendix A, and the two and three-point functions of $\zeta$ are 
defined as
\be
\langle \zeta(t,k) \zeta(t,k')  \rangle = (2 \pi)^3 \delta(k+k') \Sigma_\zeta(t, k)\,.   
\ee
and
\be
\langle \zeta(t,k) \zeta(t,k')  \zeta(t,k'')\rangle = (2 \pi)^3 \delta(k+k+k''') \alpha_\zeta(t, k,k',k'')\,.   
\ee

\subsection{Adiabaticity}

Ideally, this evaluation of the statistics of  $\zeta$ should be performed once 
the dynamics becomes adiabatic and all isocurvature modes have decayed, 
so that $\zeta$ cannot evolve any further \cite{Rigopoulos:2003ak,Lyth:2004gb}. Such a decay could occur during inflation if the mass in the 
direction orthogonal to the direction of travel becomes much heavier than the Hubble rate during 
inflation (see, e.g.,~\cite{Elliston:2011dr}), or if thermal equilibrium is reached after inflation 
ends~\cite{Weinberg:2004kf}. In many two-field models such as the Wess-Zumino model, however, 
the evolution reaches the minimum without adiabaticity being achieved. If the fields reheat at the same time into a 
thermal equilibrium stage, then the evolution of $\zeta$ from the end of inflation until 
adiabaticity is reached can usually be neglected, and the answer at the end of inflation for $\zeta$ and its 
statistics can immediately be compared with observations. This is expected to be the case if the
Wess-Zumino inflaton field decays via conventional superpotential couplings to other, lighter fields, 
and is assumed for the results presented below. 
We stress that even for the evolution 
along the real axis in the Wess-Zumino model, as well as the evolutions we probe below, the 
isocurvature mode orthogonal to the direction of travel does not decay before the fields 
reach the minimum, and the statistics of the $\zeta$ could evolve after inflation unless 
the two fields reheat together -- so simultaneous reheating is an implicit assumption both in the 
work below and in~\cite{Croon:2013ana}.

The other possibility would be ``curvaton-type" behaviour \cite{Linde:1996gt,Moroi:2001ct,Enqvist:2001zp,Lyth:2001nq, Meyers:2013gua, us2},
where one field reheats significantly before the other, and 
the energy density of radiation from the reheated field redshifts 
more rapidly than energy density of the oscillating 
field, leading to a strong evolution in $\zeta$ after inflation ends.
This could be the case, e.g., in models with multiple superfields as studied in~\cite{Ellis:2013iea}.

\subsection{Summary of numerical method}

For the study in this paper, we set up a numerical code that solves the background-field equations 
for the scalar-field cosmology, given below, 
and the transport equation (\ref{transport}) for the two-point function from vacuum initial conditions 
for the $k$ mode that cross the horizon roughly $50$ e-folds before the end of inflation, together with 
a few neighbouring modes. The code starts the evolution of the 
two-point function roughly $5$ e-folds before the $k$ mode of interest crosses the 
horizon, where assuming Bunch-Davis initial conditions is an excellent approximation. It also solves for the evolution  of the 
equilateral three-point function for this $k$ mode from horizon crossing onwards using Eq.~(\ref{transport2}).
The choice of this number of e-folds is a representative, though
relatively arbitrary choice, and we discuss later the sensitivity of our results to this assumption.
By comparing the amplitude $\Sigma_\zeta(k)$ to the 
amplitude of the two-point correlation function of 
gravitational waves, we can calculate $r$, then by comparing the amplitude of $\Sigma_\zeta(k)$
for two neighbouring $k$-modes and forming a finite difference 
approximation we can calculate $n_s = {\rm d} \ln \Sigma_\zeta(k) / {\rm d} \ln k  +  4$. 
For completeness, in a similar way we also 
calculate the the running $\alpha = {\rm d} n_s / {\rm d} \ln k  $. Finally we calculate the local $\fnl$ 
parameter which is given by the expression $\fnl =  \frac{5}{18} \alpha_\zeta(k,k,k)/\Sigma_\zeta(k)^2$.

We could of course also output the 
full spectrum, or bispectrum. In this study, however, we follow common practice and 
utilise the parameters summarised above to immediately compare the model 
to constraints derived from data.

\section{Analysis of the Wess-Zumino Model}
\label{model}

Thus far we have reviewed general techniques. 
In the following, we specialize to one particular example of 
multi-field inflation, which is of interest in its own right as well as a
representative of a broad class of models,
and explore the inflationary possibilities of the Wess-Zumino model~\cite{Croon:2013ana},
whose scalar potential $V$ is obtained from the superpotential $W$: 
\be 
W = \frac{\mu}{2} \Phi^2 - \frac{\lambda}{3} \Phi^3 ~~, ~~ V= \left |\frac{\partial W}{\partial \Phi} \right |^2 \, .
\label{WZ}
\ee
One may write the complex scalar field $\Phi = \frac{1}{\sqrt{2}}\phi \exp(i \theta)$, as in \cite{Croon:2013ana}, in which
case the scalar potential takes the form
\be
V=A\left( \phi^4 - 2 \cos(\theta) v\phi^3 + v^2 \phi^2 \right)\,.
\label{CroonV}
\ee
which reduces to the hilltop form $V = A \phi^2 (v - \phi)^2$ when $\theta = 0$.

\begin{figure}[htb]
\centering
{\includegraphics[width=0.65\textwidth]{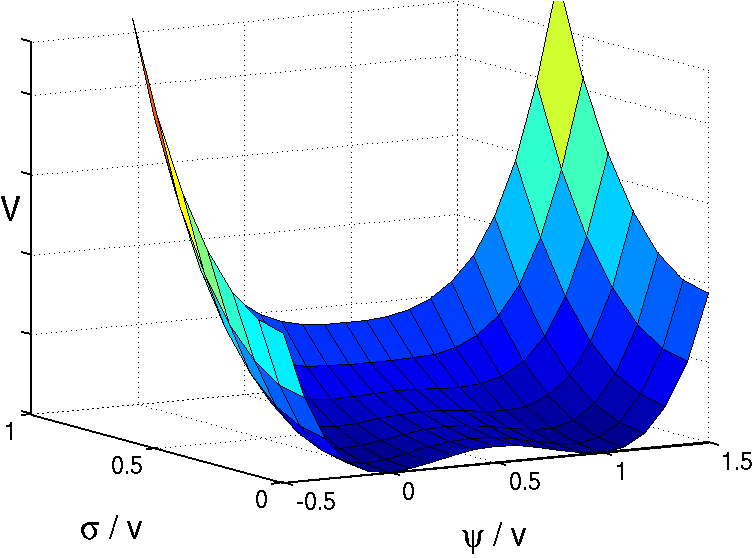}}
\caption{The Wess-Zumino potential $V$ (\protect\ref{WZpot})
as a function of the canonically-normalized real and imaginary
parts $(\psi, \sigma)$ of the complex scalar field $\Phi$ in the Wess-Zumino model (\protect\ref{WZ}),
for the choices $v=10\Mpl$ and $A=1$.}
\label{potential}
\end{figure}

However, for the purpose of our two-field study we prefer to work with canonically-normalized fields $\psi$ and $\sigma$, 
which are the real and imaginary parts of $\Phi$, i.e., $\Phi = \frac{1}{\sqrt{2}}(\psi + i \sigma)$. 
Using $\phi^2 = \psi^2 + \sigma^2$ and $\psi = \phi \cos(\theta)$, we can rewrite the potential as
\be
V = A \left ( ( \psi^2 + \sigma^2)^2 -2 v \psi(\psi^2 + \sigma^2) + v^2 (\psi^2 +\sigma^2) \right )\,.
\label{WZpot}
\ee
A visualization of the potential in this representation is shown in
Fig.~\ref{potential} for $v=10\Mpl$ and $A=1$: as well as being reflection-symmetric about $\psi=v/2 = 5\Mpl$
as displayed, we recall that the potential is also reflection-symmetric about $\sigma=0$. 

We assume canonical kinetic terms, as in the original renormalizable, globally-supersym- metric
Wess-Zumino model, in which case the effective field-space metric is flat, i.e., the K\"ahler potential 
is minimal. Therefore the background equations are
\begin{eqnarray}
H^2 &=& \frac{1}{3 \Mpl^2} \left ( \frac{\dot \varphi_a\dot \varphi_a}{2}+V \right) \,,\\
\ddot\varphi_a &=& -3 H \dot{\varphi}_a -V_{,a}\,,
\end{eqnarray}
and the perturbation equations are of the form given in (\ref{pertEvolve}), 
where we identify $\varphi_a = \{\psi, \sigma\}$.

\subsection{Model Predictions for Cosmological Observables}

In a generic two-field model, one may obtain any particular number of 
e-folds of inflation by choosing initial conditions on a one-dimensional contour in field 
space.
In the present case we are interested in the Fourier modes of 
perturbations that cross the cosmological horizon roughly 
$50$ e-folds before inflation ends.
Evolutions that pass through different locations along the $50$ e-fold 
contour can give rise to different observational predictions, and the values of the 
observational parameters that follow from different locations on this 
surface represent the exhaustive range of values the model can give yield
for any particular choice of model parameters.

To determine whether the Wess-Zumino model can be compatible with 
current observational constraints for a particular choice of the model parameter $v$, therefore, we 
first perform a numerical search to find points close to the the contour that gives
$50$ e-folds of inflation for some representative values of $v$. We allow points 
that lead to $N=50 \pm 0.2$.
We then use the tools described above to calculate the observables $n_s$, $r$ and $\fnl$ 
that correspond to each position found. We aim to populate the contour sufficiently 
densely so that its shape is apparent, though the precise points found depend on the details of 
the search performed.
We note that the value of $A$ does not affect the 
number of e-folds, $n_s$, $r$ or $\fnl$.
Hence, $A$ can be fixed independently at every position on the $50$ e-fold contour,
so that the amplitude of $\zeta$ is in agreement with observations. 

This analysis is performed in \S~\ref{range}. As we will see 
it is possible for the model to produce a range of signatures, and one
might wonder whether all these possible signatures can actually be realized if, for example, 
inflation begins at some much higher energy scale so that many 
more than $50$ e-folds occur. In this case, although only the last $50$ 
e-folds play a role in the generation of observational 
signatures, the evolution during the 
preceding e-folds determines where on the $50$ e-fold 
contour the inflationary trajectory passes. In \S~\ref{stats}, therefore, we look at the model 
in this alternative way, beginning the evolution on a constant-energy-density surface 
inside the eternal inflation boundary at large field values. Moreover, we 
follow Frazer~\cite{Frazer:2013zoa}, and ask not only what observational 
signatures can be realized from these initial conditions, but also what are the most likely signatures, 
by producing a probability distribution of the resulting values of $n_s$ and $r$ from an
assumed uniform prior of the likelihood of initial conditions on the constant-energy-density 
surface.

\begin{figure}[h]
\centering
{
\includegraphics[width=0.54\textwidth]{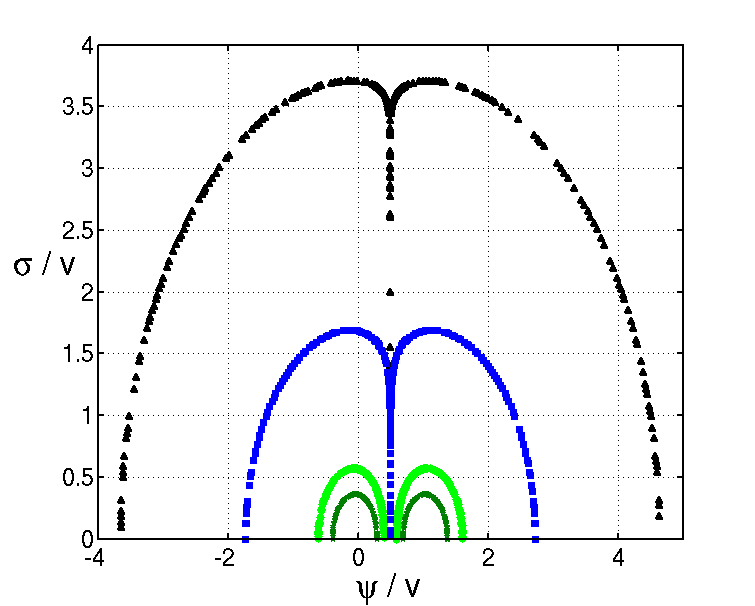}
\mbox{\includegraphics[width=0.44\textwidth]{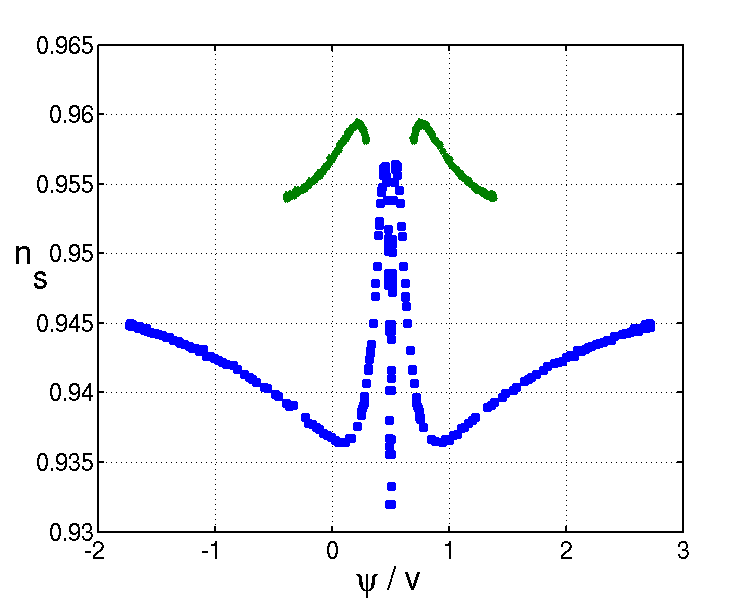}
\includegraphics[width=0.44\textwidth]{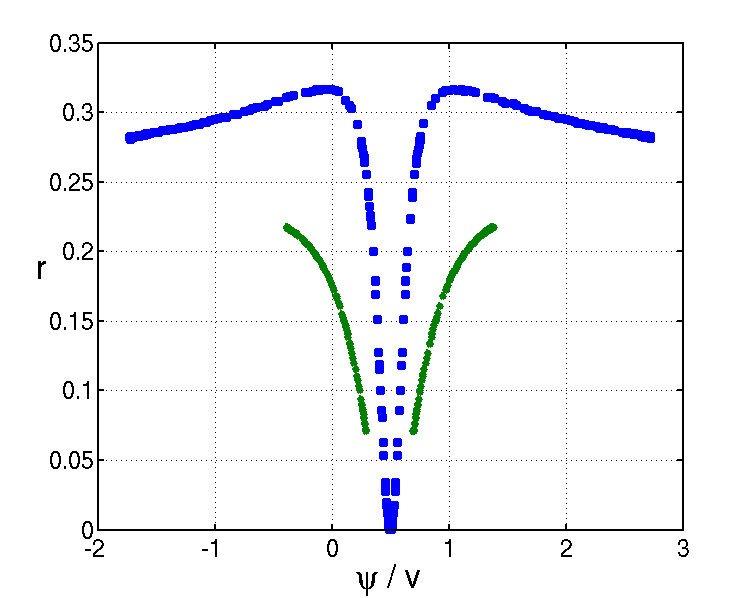} }
\mbox{\includegraphics[width=0.44\textwidth]{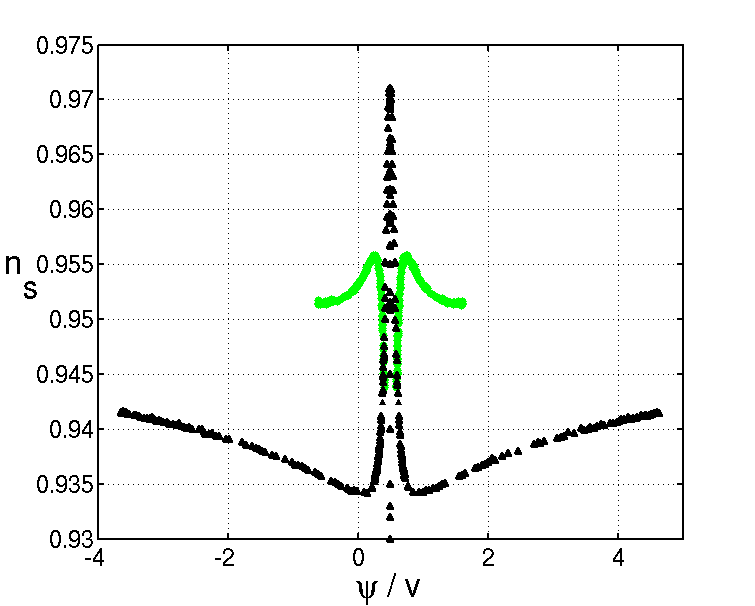}
\includegraphics[width=0.44\textwidth]{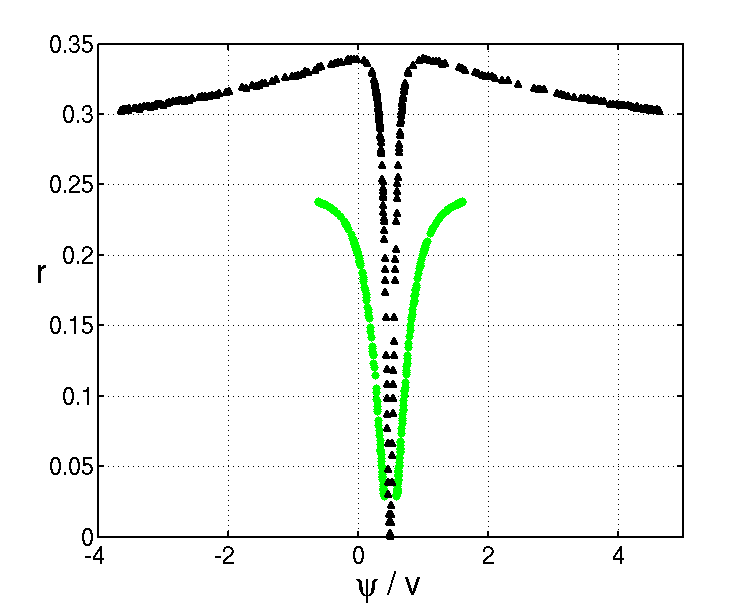} }}
\caption{Points on the $50$ e-fold contour are indicated with dark green 
stars for $v = 40\Mpl$, with light green 
circles for $v=26\Mpl$, with 
 blue squares for $v=10\Mpl$, 
and with black triangles for $v=5\Mpl$ (upper panel), and the corresponding values
of the observables $n_s$ (middle left panel for $v=40\Mpl$ and $v=10\Mpl$, and lower left panel 
for $v=26\Mpl$ and $=5\Mpl$) and $r$ (middle right panel for $v=40\Mpl$ and $v=10\Mpl$, and lower left panel 
for $v=26\Mpl$ and $=5\Mpl$). The $n_s$ and $r$ plots are 
split between the middle and lower panels for visual clarity. In each case, the 
lines probed by the marked points are more extended the smaller the value of $v$.  
The field axes are chosen to be $(\hat \psi \equiv \psi/v, \hat \sigma \equiv \sigma/v)$.}
\label{results1}
\end{figure}

\section{Results}
\label{results}
\subsection{Predictions for $n_s$ and $r$}
\label{range}

It was found in~\cite{Croon:2013ana} that, if the initial condition was chosen so that $\sigma = 0$,
successful predictions for inflationary observables could be obtained for $v \sim 40\Mpl$ to $60\Mpl$.
Accordingly, we first explore how the results of~\cite{Croon:2013ana} can be generalized if one
chooses $v = 40\Mpl$ but allows initial conditions with $\sigma \ne 0$. Later we explore the
possibilities if $\sigma \ne 0$ in the cases that $v=26\Mpl$, $v = 10\Mpl$, and $v=5\Mpl$, 
values that were not consistent with the data if $\sigma = 0$. The results for $v=40\Mpl$ 
and $v=26\Mpl$ can be compared with 
Table~1 of~\cite{Croon:2013ana}

Positions on the $50$ e-fold contour for $v = 40\Mpl$ are shown as the set of dark green star 
shaped points in the upper panel of Fig.~\ref{results1},
where the field axes are chosen to be $(\hat \psi \equiv \psi/v, \hat \sigma \equiv \sigma/v)$. The points 
run into one another in this case, but the contour can be identified as the inner most one on the figure.
The contour takes the form of twin rings around the two minima of the potential, which intersect
the real axis at $\psi \sim 12\Mpl$, $\hat \psi \sim 0.3$,
(the solution found in~\cite{Croon:2013ana}), at
$\psi \sim -15.5\Mpl$, $\hat \psi \sim - 0.39$, and at an equivalent pair of points that are their mirror images under reflection in the symmetry axis. 
The middle left panel of Fig.~\ref{results1} shows the $n_s$ predictions for the points 
along the 50 e-fold contour
for $v = 40\Mpl$, and we see that all the points are consistent with the Planck constraint 
$n_s = 0.9603 \pm 0.0073$ at the 68\% CL. 
However, we see in the middle right panel that the predictions for the tensor-to-scalar ratio, $r$, are in general less successful. Indeed, the only points along the left ring in the upper panel of
Fig.~\ref{results1} that yield $r < 0.08$, and hence are consistent with the Planck data at the 68\% CL, are
those with $\psi > 10\Mpl$. We see from the upper panel of Fig.~\ref{results1} that these points also have $|\sigma| < 10\Mpl$.
These points and their mirror images along the right ring in the upper panel of Fig.~\ref{results1}
generalize the solution found in~\cite{Croon:2013ana} for $v = 40\Mpl$, which had $\sigma = 0$. They 
represent trajectories evolving on the concave, hill, region of the potential, with most of the 
evolution in the $\psi$ direction. 

\begin{figure}[h]\centering
{\includegraphics[width=0.7\textwidth]{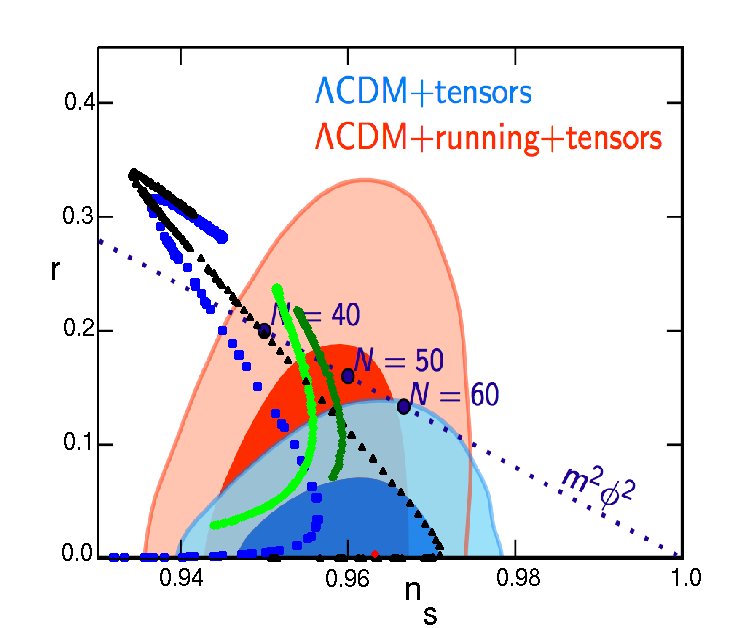}}
\caption{The 50 e-fold results for $n_s$ and $r$ superimposed on the Planck constraints taken from Fig.~23 of Ref~\cite{Ade:2013zuv}. 
The $v =40\Mpl$ results are
represented by the dark green stars (this is the smallest arc on the diagram), the $v=26\Mpl$ results 
by the light green circles, the $v=10\Mpl$ results by the blue squares, and the $v=5\Mpl$ results by the black 
triangles. Once again the lines probed by our results are more extended the smaller the value of $v$.
For comparison, we also display the corresponding result for the Starobinsky model of $R^2$ 
inflation~\cite{starobinsky} with $50$ e-folds (red diamond).}
\label{results2}
\end{figure}

Our generalized results for $v = 40\Mpl$ and their compatibility with Planck constraints are summarized in Fig.~\ref{results2} which is discussed further below.

Our second example is $v=10\Mpl$, marked on Fig.~\ref{results1} by 
blue squares in the top and middle panels. In this case the restriction to the real axis
discussed in~\cite{Croon:2013ana} does not yield successful inflation in the hilltop
region between the twin minima. For $\sigma = 0$,  
one can obtain $50$ e-folds by choosing initial conditions extremely close to the top of the 
hill at $\psi = 5\Mpl$, $\hat \psi \simeq 0.5$ (where N diverges to infinity), 
but for this case $n_s$ is much too small, 
in fact it is off the bottom of the plot in the middle left panel of Fig.~\ref{results1}. However, 
$50$ e-folds also results from positions 
along an arc extending from $(\psi, \sigma) \sim (5, 0)$ through points
with $\sigma \ne 0$ down to $(\psi, \sigma) \sim (-17, 0)$ (or along the corresponding mirror contour). 
We see in the middle left panel that, for points corresponding to the left ring in the upper panel, 
suitable values of
$n_s$ within the $n_s = 0.9603 \pm 0.0073$ range are found in the range $4.2\Mpl < \psi < 4.8\Mpl$, $ 0.42 <\psi < 0.48$.
In these cases, as seen in the middle right panel of Fig.~\ref{results1},
the corresponding values of $r$ are small, namely in the range
$0.07 > r > 0.007$. The ``good" inflationary trajectories start close to the
minimum of the valley seen in Fig.~\ref{potential} at $\psi = 5\Mpl$ and large $\sigma$, 
roll along the $\sigma$ direction, and then fall
off after the valley has turned into a ridge and roll to one of the two equivalent minima.
We re-emphasize that the model seemed to be completely ruled out for $v=5\Mpl$,
as long as one considered evolution only in the real direction, but that
initial conditions consistent with observations can be found if one allows
inflationary trajectories that start elsewhere in the two-field space.

These results and their compatibility with Planck constraints are summarized in Fig~\ref{results2}.
We see that some initial conditions on the 50 e-fold contour for $v = 10\Mpl$ (blue squares) 
yield values of $(n_s, r)$ within
the Planck 68\% CL region, both for the joint 
distribution including tensors and the joint distribution including  
tensors and running. On the other hand, the line of $v = 40\Mpl$ points (dark green stars) 
approaches but does not quite enter this region for the distribution including 
tensors alone
for 50 e-folds.

In order to determine the how representative the values of $v$ we have considered thus far, finally 
we consider two further values, $v=26\Mpl$ and $v=5\Mpl$. The $v=26\Mpl$ results are represented by 
light green circles in the top and bottom panels of Fig.~\ref{results1}. For the hilltop $\sigma=0$ trajectory, 
although $r$ is small, $r\sim0.03$,  $n_s$ lies outside the $n_s = 0.9603 \pm 0.0073$ 
region. When other evolutions 
are considered, the $v=26\Mpl$ case exhibits behaviour intermediate between the $v=10\Mpl$ and $v=40\Mpl$ examples. Initial conditions with $-0.1\Mpl<\psi<9.1\Mpl$ provide values of $n_s$ in 
the range $n_s = 0.9603 \pm 0.0073$ but for this range we find $r>0.07$. From Fig.~\ref{results2}, 
one can see that no initial conditions lead to a value of $(n_s, r)$ within the 
the Planck 68\% CL region for the joint distribution without running, but points 
are found within the 68\% CL region when  
tensors and running are included. 
The $v=5\Mpl$ case, represented in the top and bottom 
panels of Fig.~\ref{results1} by black triangles, exhibits similar features to the $v=10\Mpl$ 
case, however, a broader range of $n_s$ and 
$r$ is possible, as is also seen in Fig.~\ref{results2}. As is seen from these 
figures, this case can be consistent with observations 
for similar valley/ridge 
trajectories to the ones described for the $v=10\Mpl$ example, despite being completely ruled out 
if only evolution in the real field direction were to be considered. 

Some further comments are in order. 
First we note that for all the points shown in Fig.~\ref{results2}
we find that the running of the spectral index is 
within the range $ -0.003 < \alpha < 0$. The magnitude of this negative running is 
smaller than that preferred by the Planck results at  
68\% CL level (see Fig.~23 of Ref.~\cite{Ade:2013zuv}.), but well within the $2\sigma$ results 
which are consistent with zero running.
Secondly, it is interesting to ask how the curves on Fig.~\ref{results2} change if we select 
a different number of e-folds from the scales of interest crossing the horizon until the end of inflation.
This true number is determined by the post inflationary dynamics such as the energy scale of 
reheating, and is generally taken to be between $40$ and $60$. Although we have 
not performed an exhaustive study, the rule of thumb is that the main change for 
curves on Fig.~\ref{results2} is that they are   
shifted to the right and down with increasing number of e-folds, although the shapes are also slightly 
distorted. We choose $50$ as a representative number for the allowed range, and to allow comparison with 
Ref.~\cite{Croon:2013ana}.
Finally, we stress that in all the plots presented thus far, 
no inference should be drawn from the number of points populating 
different parts of the curve in Fig.~\ref{results1} or Fig.~\ref{results2}, as this is a consequence 
of our scanning strategy.

Before closing this Section, we comment on the
comparison of our two-field Wess-Zumino  supersymmetric model for inflation
with the Starobinsky model~\cite{starobinsky}, whose excellent agreement with the Planck data (see Fig.~\ref{results2}) 
has recently prompted a plethora of interesting works finding Starobinsky-type potentials in 
supergravity models~\cite{starosugra}. As we see in Fig.~\ref{results2}, 
the two-field Wess-Zumino model for $N=50$ e-foldings and  $v=5\Mpl$ or $v=10\Mpl$
can yield results similar to the Starobinsky model, for 
particular initial conditions close to the $\sigma=0$ axis. 

\begin{figure}[h]
\centering
{\includegraphics[width=0.7\textwidth]{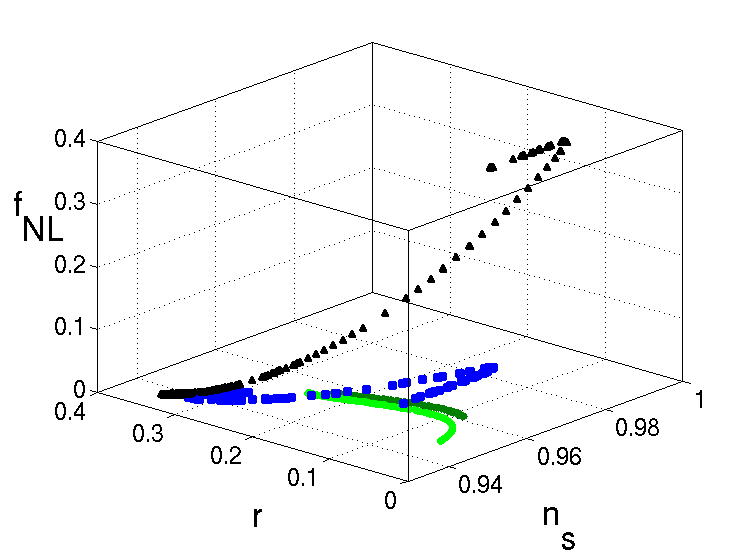}
\mbox{\includegraphics[width=0.54\textwidth]{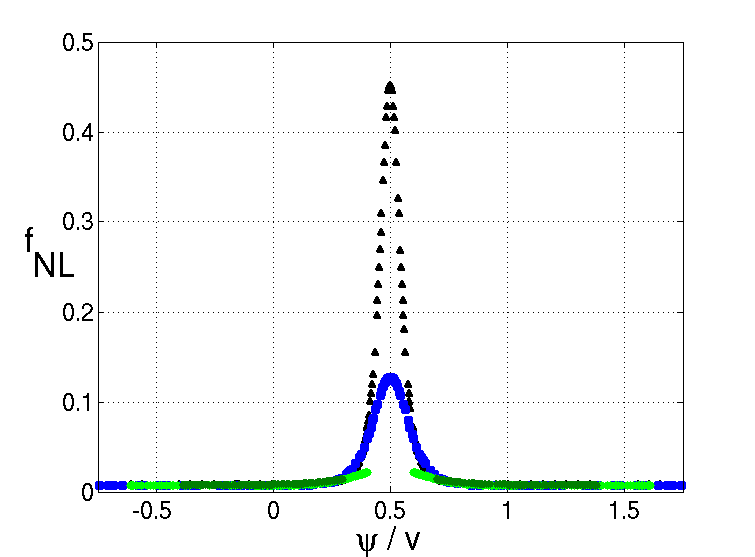}
\includegraphics[width=0.5\textwidth]{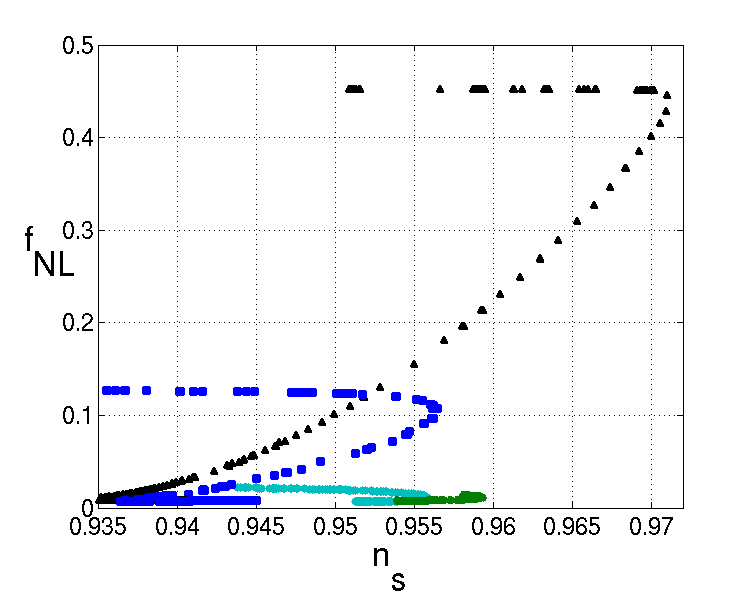}}}
\caption{The 50 e-fold results for $n_s$, $r$ and $\fnl$ (upper panel). And for clarity the plots for $\fnl$ against 
$\psi/v$ and $\fnl$ against $n_s$ (lower left and lower right panels respectively). As before 
$v = 40\Mpl$ results are marked with dark green stars, $v=26\Mpl$  with light green 
circles, $v=10\Mpl$ with 
blue squares and $v=5\Mpl$ with black triangles. }
\label{fnl}
\end{figure}

\subsection{Predictions for $\fnl$}
\label{fnl}

We now consider the Wess-Zumino model predictions for
$\fnl$. In simple models of inflation, a value of $\fnl$ greater than the order of magnitude of the 
slow-roll parameters, i.e., $\gtrsim 1/N^*$ where $N^*$ is the observable number of e-folds, requires 
rather specific evolutions~\cite{Byrnes:2008wi, Elliston:2011dr, Elliston:2012wm}. On the other hand, 
ridge-type evolutions such as those discussed above are known to play a r\^ole in producing 
larger values of $\fnl$. 
We have explored the values of $\fnl$ that are produced in the cases discussed above
using the calculational method described earlier, with the results summarized in Fig~\ref{fnl}.  
For the $v=40\Mpl$ and $v=26\Mpl$ cases, we can see that $\fnl$ is never significantly enhanced
above the slow-roll value. On the other hand, for $v=10\Mpl$ and $v=5\Mpl$, 
one can see that the ``ridge" trajectories that leads to consistent 
$n_s$ and $r$ values do indeed lead to values of $\fnl$ that are enhanced by over an order
of magnitude, and that the largest enhancements we find are for the small values of $r$ that are
favoured by the Planck measurements.  For $v=10\Mpl$ we find that $\fnl \lesssim 0.13$, and for 
$v=5\Mpl$ $\fnl\lesssim 0.45$,  
too small to be probed by present experiments. These number are 
never-the-less an interesting signature of ridge trajectories, and are consistent 
with the ``ridge'' estimate 
of $\fnl \sim -( 5 V_{,\sigma \sigma})/(6 V)$ \cite{Kim:2010ud, Elliston:2011dr, Mulryne:2011ni}, despite this 
being derived for ``separable potentials'', and not strictly being applicable to 
the current setting. For the Wess-Zumino model this estimate 
yields $\fnl \sim 13.3/v^2$, and hence $\fnl \sim 0.13$ for $v=10\Mpl$ and $\fnl \sim 0.53$ for $v=5\Mpl$ in reasonable agreement with the results presented above.

\begin{figure}[h]
\centering
{\includegraphics[width=0.7\textwidth]{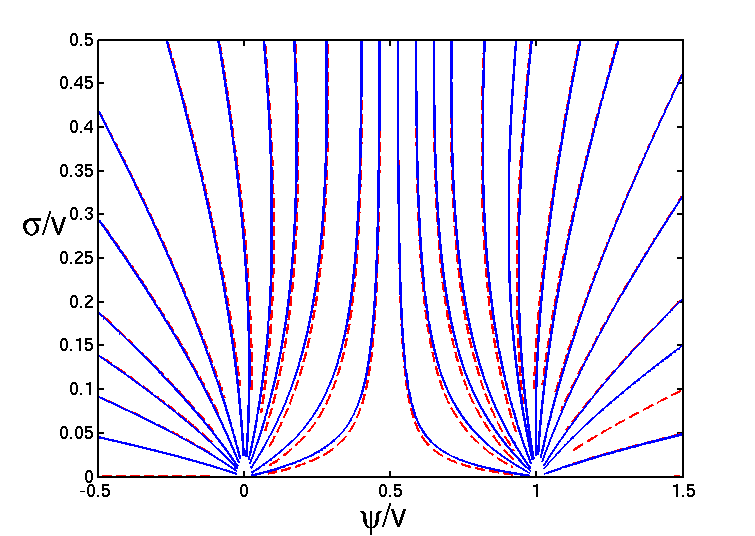}}
\caption{Representative trajectories from a particular surface of constant energy density fixed 
outside of the the visible area. The blue solid lines are trajectories for the $v=40\Mpl$ case 
and the red dotted lines are for the $v=10\Mpl$ case. We see that while slow-roll is a good 
approximation these lines are almost identical when plotted on the rescaled $(\psi/v , \sigma/v)$ 
axis, but diverge as slow-roll ends. Inflation also ends in different 
positions for the two cases. }
\label{trajectories}
\end{figure}

\subsection{Statistical Predictions}
\label{stats}

Finally, we consider the probability distributions of possible inflationary predictions
in the Wess-Zumino model assuming a uniform prior for initial conditions along
a contour of constant potential energy density in a similar manner 
to Refs.~\cite{Frazer:2013zoa, Easther:2013rva}.

At much larger field values than are probed by the last $50$ e-folds of evolution 
for any inflationary trajectory, the Wess-Zumino potential has the approximate form 
$V \sim A (\psi^2 +\sigma^2)^2 $. Thus, trajectories that start at large 
field values feel initially a potential that is approximately quartic in form, and the
potential is rotationally symmetric.
In this regime, constant-energy-density surfaces are therefore circles 
defined by $\psi^2 + \sigma^2={\rm Const}$. By picking 
a particular value of the constant we choose a particular energy. In order to probe the 
model statistically, we parametrize this circle using 
polar coordinates, fix the initial energy and hence the initial 
radial coordinate, and pick initial angular coordinate values by drawing from
a uniform distribution. This prior expresses maximal ignorance about 
the origin of trajectories.

The trajectories that arise from such a surface can be seen 
in Fig.~\ref{trajectories}. The trajectories are initially straight lines that
intersect successive circular uniform density surfaces. Therefore, as long as the 
field values are sufficiently large for the potential to be rotationally 
symmetric and a uniform prior is employed, the results we present 
are independent of the energy of the surface on which we choose 
to fix initial conditions. This would of course not be true if an alternative surface or prior were 
chosen, or the trajectories were initiated at a much lower energy scale. As the 
trajectories approach the region which contains the minima they begin to 
curve and eventually fall into one of the two minima as seen in Fig.~\ref{trajectories}.

\begin{figure}[h]
\centering
{\includegraphics[width=0.8\textwidth]{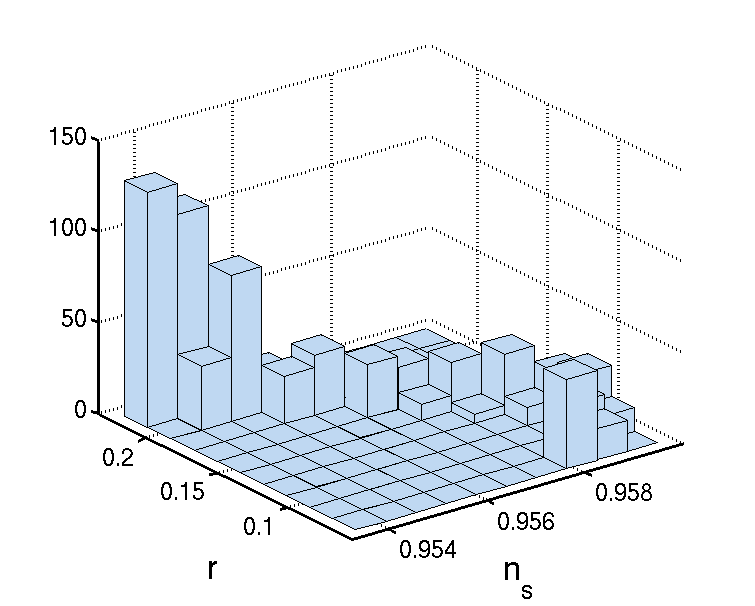}
\includegraphics[width=0.67\textwidth]{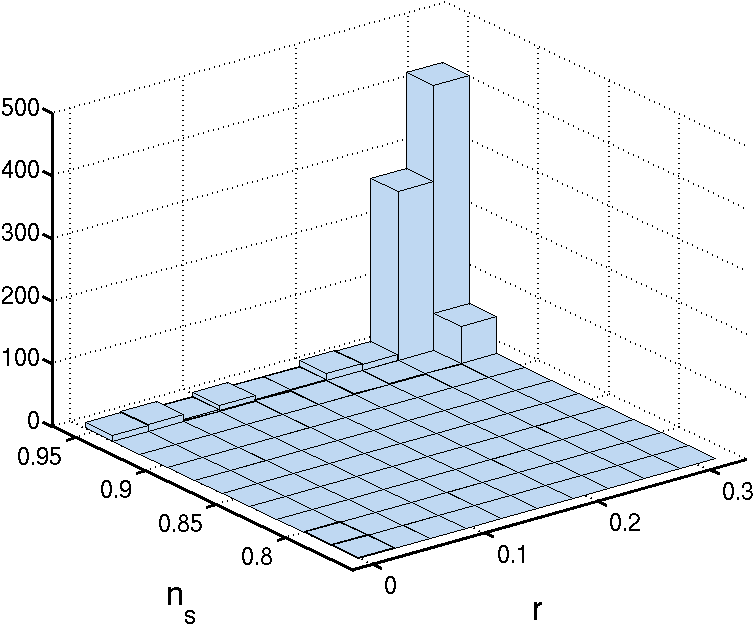}}
\caption{The statistical distributions of $n_s$ and $r$ for the $v=40\Mpl$ (upper panel) and $v=10\Mpl$ (lower panel) cases. The results 
presented are not normalised, and so are histograms with the vertical axis referring to the number 
of points in each bin. }
\label{stats40}
\end{figure}

The resultant distributions for $n_s$ and $r$ for the $v=40\Mpl$ and $v=10\Mpl$ cases 
are plotted in Fig.~\ref{stats40}, respectively, where roughly $1000$ 
initial conditions are drawn from each of the prior distributions (the results are presented in 
histogram form, and so the vertical axis represents the number of points in each bin, rather than 
the normalised probability distribution).
One can see in both cases that the bulk of the probability distribution corresponds to values of $r$ disfavoured by Planck.
However, in the $v=40\Mpl$ case there is significant weight close to the preferred region.

It is an interesting question how to interpret these results. First we should note that we have effectively
marginalised over the amplitude of the scalar power spectrum, since we have fixed this 
by implicitly normalising $A$. This is true also of previous statistical 
studies \cite{Frazer:2013zoa, Easther:2013rva}, and in reality 
there is a distribution of amplitudes as well as of $n_s$ and $r$ given 
fixed model parameters.  This makes 
it difficult to directly interpret the plotted distributions. Moreover, even if we were to take the 
distributions 
at face value, it is still an open question when 
to prefer one case over another. For example, comparing the $v=10\Mpl$ and $v=40\Mpl$ cases
it is  clear more of the probability in the $v=40\Mpl$ case is close to the  observationally preferred 
values of $n_s$ and $r$.  
However, the $v=10 \Mpl$ case has a small probability
to get right in the middle of the preferred $(n_s, r)$ region. We might consider therefore not weighting 
all initial conditions equally, but preferring those that get closer to the the preferred observational values 
of $n_s$ and $r$, and thus deriving some ultimate measure of the goodness of the model. Though that is 
beyond the scope of the present work. On the other hand, if we simply required the majority of the 
probability as plotted to be within the preferred $(n_s, r)$ region, it is clear that neither value 
of $v$ could be considered consistent with observations.
 

Before concluding this Section, we re-emphasize the point that, in this model,
trajectories that begin their evolution in the quartic regime of the potential may plausibly find their 
way into the concave region for the requisite number of e-folds, as exemplified by the $v = 40$ 
case discussed above -- where a minority but significant amount of the probability distribution 
corresponds to evolutions on the 
concave region. This observation is interesting in 
light of recent criticisms of inflation~\cite{Ijjas:2013vea}. Analyzing a single-field model of the form of
(\ref{CroonV}) with $\theta = 0$, it was argued in~\cite{Ijjas:2013vea} that hilltop-like 
regimes preferred by Planck data Planck-compatible hilltop-like inflation would be exponentially unlikely.
The reasoning was that they would require an energy density $\ll M_P^4$, whereas
the primordial universe is likely to have had a much higher energy density, e.g., ${\cal O}(M_P^4)$,
in which case chaotic inflation with large field values and hence unacceptably large energy density would be exponentially more likely. 
However, the Wess-Zumino case studied here, which allows $\theta \ne 0$ in (\ref{CroonV}),
is a counter-example showing that a realistic two-field dynamical model may
allow as generic possibilities trajectories that find their way into a region preferred by cosmological observations, even
if they originate at extremely high energy densities~\footnote{This argument of~\cite{Ijjas:2013vea} has also been criticized
recently in \S~V of~\cite{Guth:2013sya}.}.

Finally we note that in considering the trajectories prior to the last $50$ e-folds, we have only 
considered the classical trajectory and not stochastic fluctuations about that path. While 
those fluctuations will be sub-dominant in each Hubble time 
outside of the eternal inflationary regime, the will accumulate over time and might 
still affect the statistics we have presented, and their effect on the statistics is an open question.

\section{Conclusions}
\label{conclusions}

We have found in this exploration of the simplest Wess-Zumino model of inflation
with real and imaginary field components several features of interest, that may 
also  
be present for more elaborate multi-field models. The introduction of the second (imaginary)
degree of freedom opens up new possibilities for successful inflation that were not
discussed in~\cite{Croon:2013ana}. For example, we found that inflation is possible
for values of the field scale $v$ that are much smaller than in the case where only
the real part of the Wess-Zumino field is considered. We also found that much smaller
values of $r$ could be obtained than in the previous simplified single-field ``hill-top" 
treatment of this model. We also showed that $\fnl$ could be much larger than usually
found in slow-roll models, although the largest values we found were still too small to
be observable in current experiments.

We also made an estimate of the statistical probability distribution of possible
inflationary parameters, assuming pre-inflationary initial conditions corresponding
to a uniform prior on a contour of constant energy density in the model parameter space.
Although the probability distribution is not maximized in the regions of parameter space
favoured by the Planck data, neither do these regions correspond to severe fine-tuning 
at least for the $v=40\Mpl$ case.
The fact that one can find suitable inflationary trajectories starting from a high
pre-inflationary energy density weakens one of the criticisms of single-field inflation made
recently in~\cite{Ijjas:2013vea}, where it was argued that hill-top scenarios were unlikely
to be realized.

We regard as very promising this exploration of the new inflationary possibilities that open up in two-field
models. Although the Wess-Zumino model has an excellent theoretical pedigree, there are many
proposals for more elaborate multi-field models that embody more attractive features. Staying within
the supersymmetric framework, for example, it is desirable to incorporate the modifications to
the effective potential induced by supergravity, and interesting to consider models in which local
supersymmetry is broken dynamically~\cite{locadyna}. The analysis of this paper indicates that these models
might exhibit interesting novel features when their full complexity is considered.

\bibliographystyle{JHEPmodplain}
\footnotesize
\bibliography{wess}

\section*{Appendix A}
In this appendix, in order that 
readers can reproduce our results, we summarise the super-horizon 
equations of motion for the scalar field fluctuations 
at second order, needed for Eq.~(\ref{transport2}) which we use to  
calculate $\fnl$, and we also provide the relations we use to connect 
the field fluctuations to $\zeta$. The methods 
by which these are calculated are discussed extensively 
elsewhere \cite{otherTransport}\footnote{We recall that the linear cosmological parameters, $(n_s, r, \alpha)$ 
are calculated using the full first order 
$k$ dependent equations of motion for the two-point 
function from 
vacuum initial conditions, given in the main body of the text.
But that the three point function is 
calculated assuming it is zero at  
horizon crossing, and using only the super-horizon equations of 
motion.}

We can write the second order equations of motion 
for the scalar field fluctuations, at second order in 
perturbation theory, as a set of first order equations of the compact form
\be
\frac{{\rm d} x_\alpha}{{\rm d} N} = u_{\alpha \beta} x_{\beta} + \frac{1}{2} u_{\alpha \beta \gamma} x_{\beta}x_{\gamma}
\ee
where for convenience we use 
$N$ as the time variable, the Greek indices run over all fields and 
field momenta perturbations, and where 
$x_\alpha$ represents
$(\delta \phi_a , \delta  \phi'_{b})$. 
The  $u$ matrices are 
\begin{eqnarray}
u_{a  b} &=& 0 \,, \nonumber \\  
u_{a \bar b} &=& \delta_{a b} \,,\nonumber \\  
u_{\bar a b} &=& - M^2_{a b}/H^2  \,, \nonumber \\
u_{\bar a \bar b} &=& (\epsilon-3) \delta_{a b} \,, \nonumber \\
u_{ a  b c} &=& 0 \,, \nonumber \\
u_{ a  \bar b c} &=& 0 \,, \nonumber \\
u_{ a  \bar b \bar c} &=& 0 \,, \nonumber \\
u_{ \bar a  b c} &=& -\frac{V_{,a b c}}{H^2} + \left [\frac{V_{,a b} V_{,c} }{H^4 \Mpl^2 (3-\epsilon)} + (b \rightarrow c) \right] + 
\frac{V_{,b c} V_{,a} }{H^4 \Mpl^2 (3-\epsilon)}  - 
\frac{2 V_{,a}V_{,b}V_{,c} }{H^6 \Mpl^4 (3-\epsilon)^2}    \,, \nonumber \\
u_{ \bar a  \bar b c} &=&   -\frac{V_{,a} V_{,c} \varphi'_b}{H^4 \Mpl^4 (3-\epsilon)^2} + \frac{V_{,a c} \varphi'_b}{H^2 \Mpl^2(3-\epsilon)}   \,, \nonumber \\
u_{ \bar a  \bar b \bar c} &=&   \frac{\varphi'_a}{\Mpl^2} \delta_{b c} + \left [\frac{\varphi'_c}{\Mpl^2} \delta_{a b} + (b \rightarrow c) \right ] - \frac{V_{,a} \varphi'_b \varphi'_c}{H^2\Mpl^4 (3-\epsilon)^2} + \frac{V_{,a} \delta_{b c} }{H^2 \Mpl^2 (3-\epsilon)}  
\,, 
\end{eqnarray}
where the bar indicates field momentum indices 
(indices to be contracted with $\delta \varphi_a'$) and no 
bar represents field indices, 
a dash indicates differentiation with respect to $N$, and  $M^2_{a b}$ is defined 
below Eq.~(\ref{pertEvolve}) 
(note that after the change in time variable is taken into account, comparing the 
first order $u$ matrices with Eq.~(\ref{pertEvolve}), the only difference is the missing 
$k^2$ term which we neglect on super-horizon scales).
These $u$ matrices are used in Eq.~(\ref{transport2}).

To convert to $\zeta$, Eqs.~(\ref{zeta1})-(\ref{zeta2}), 
we use the follow $N$ coeficients
\begin{eqnarray}
N_{\alpha} &=& -\frac{1}{2 \dot H} \frac{\partial H^2}{ \partial X_\alpha}  \,, \nonumber \\ 
N_{\alpha \beta}&=&-\frac{1}{2 \dot{H}} \frac{\partial^2 H^2}{ \partial X_\alpha X_\beta} - \frac{\partial}{ \partial X_{(\alpha}}\left(\frac{1}{\dot{H}}\right)
\frac{\partial H^2}{ \partial X_{\beta)}}+\frac{1}{2 \dot{H}}\frac{\partial}{\partial X_\gamma}\left(\frac{1}{2 \dot{H}}\right) \frac{\di X_\gamma}{\di N} \frac{\partial H^2}{\partial X_\alpha} \frac{\partial H^2}{X_\beta}\,,
\end{eqnarray}
where $X_\alpha$ represents $(\varphi_a, \varphi'_a)$ and where
\begin{eqnarray}
\frac{\partial H^2}{ \partial \varphi_a} &=& \frac{V_{,a}}{\Mpl^2(3-\epsilon)}\,, \nonumber \\
\frac{\partial H^2}{ \partial \varphi'_a} &=& \frac{H^2 \varphi'_a}{\Mpl^2(3-\epsilon)}\,, \nonumber \\
\frac{\partial^2 H^2}{ \partial \varphi_a \partial \varphi_b} &=&  \frac{V_{,a b}}{\Mpl^2(3-\epsilon)}\,,
\nonumber \\
\frac{\partial^2 H^2}{ \partial \varphi_a \partial \varphi'_b} &=&  \frac{V_{,a} \varphi'_b}{\Mpl^4(3-\epsilon)^2}\,,
\nonumber \\
\frac{\partial^2 H^2}{ \partial \varphi'_a \partial \varphi'_b} &=&  \frac{2H^2 \varphi'_a \varphi'_b}{\Mpl^4(3-\epsilon)^2} + 
\frac{ H^2}{\Mpl^2(3-\epsilon)}\delta_{ab}  \,,
\nonumber \\
\frac{\partial}{ \partial \varphi_a } \left(\frac{1}{\dot H} \right) &=&  \frac{-1}{\dot H H^2} \frac{\partial H^2}{ \partial \varphi_a} \,,
\nonumber \\
\frac{\partial}{ \partial \varphi'_a } \left(\frac{1}{\dot H} \right) &=&  \frac{ H^2 \varphi'_a}{\Mpl^2 {\dot H}^2} - \frac{1}{\dot H H^2} \frac{\partial H^2}{ \partial \varphi'_a} \,.
\end{eqnarray}

\section*{Acknowledgements}

We thank fellow participants in the London Cosmology Discussion Meetings
for interesting discussions on this subject.
The work of J.E. and N.E.M. was supported in part by the London Centre for Terauniverse Studies (LCTS),
using funding from the European Research Council 
via the Advanced Investigator Grant 267352. D.J.M. is supported by the 
Science and Technology Facilities Council grant ST/J001546/1 and would like 
to thanks Joseph Elliston for useful comments and discussion.


\end{document}